\documentclass[12pt, a4paper]{article}
\usepackage{jheppub}
\usepackage[english]{babel}
\usepackage{blindtext}
\usepackage{avant} 
\usepackage{amsmath,epsf,amssymb,latexsym,amsthm,setspace,array,pifont,color,hyperref,amsfonts,dsfont,cancel,braket,parskip,appendix,tikz,todonotes}
\usepackage[compat=1.1.0]{tikz-feynman}
\usetikzlibrary{positioning,arrows}
\usetikzlibrary{decorations.pathmorphing}
\usetikzlibrary{decorations.markings}
\usepackage[none]{hyphenat} 
\usepackage{graphicx} 
\usepackage{slashed, feyn, comment}
\newcommand{\bt}[1]{\textbf{#1}}
\newcommand{\da}{\dot{a}}
\newcommand{\db}{\dot{b}}

\newcommand{\cl}[1]{\mathcal{#1}}

\renewcommand{\to}{\longrightarrow}

\newcommand{\ang}[1]{\langle #1 \rangle}
\newcommand{\squ}[1]{[ #1 ]}
\newcommand{\sbra}[1]{[ #1 \vert}
\newcommand{\sket}[1]{\vert #1 ]}

\def\pd{\partial}

\def\spvecA#1;{\if;#1;\else #1\cr \expandafter \spvecA \fi}

\tikzset{
	particle/.style={thick,draw=black},
}
\tikzset{
	graviton/.style={
		double,
		decoration={snake, aspect=0.75, mirror, segment length=1.5mm},
		decorate
	}
}

\newcommand{\td}[1]{
\if\notesOn1
\todo[inline]{#1}
\fi
}

\def\notesOn{1}
\title{On-Shell Perspectives on the Massless Limit of Massive Supergravity}
\author[a,b] {Daniel J Burger}
\author[a,b] {, Nathan Moynihan}
\author[a] {\& Jeff Murugan}

\affiliation[a]{The Laboratory for Quantum Gravity \& Strings (QGASLab)}
\affiliation[b]{High Energy Physics, Cosmology \& Astrophysics Theory (HEPCAT) group ,\\ 
Department of Mathematics,\\
University Of Cape Town,\\
Private Bag, Rondebosch, 7701, South Africa}
\emailAdd{burgerjdaan@gmail.com}
\emailAdd{nathantmoynihan@gmail.com}
\emailAdd{jeff.murugan@uct.ac.za}
\abstract{Massive gravity exhibits a famous discontinuity in its 2-point linearized amplitude for t-channel scattering of gravitational sources, in the $m\to0$ limit. In essence, the source of this vDVZ discontinuity is in the failure of the zero-helicity mode to decouple in this limit. In \cite{Moynihan:2017tva}, we showed how this result could be understood in the context of modern on-shell methods and, in particular, the BCFW construction. In this article, we provide a similar on-shell perspective to the equally interesting but lesser known discontinuity first discovered by Deser, Kay and Stelle in massive supergravity.}

\begin{document}
	\sloppy
	\maketitle
\section{Introduction}
Endowing the graviton with a small but nonzero mass is an appealing idea for many reasons, the most significant of which is a credible explanation for the observed late-time acceleration of the Universe without the need to invoke exotic forms of matter and energy. However, since nothing in life is free, this approach is not without its own pathologies. Among these are a ghost mode and a, less scary but perhaps more famous, discontinuity in the 2-point function of the theory in the massless limit. The former was  resolved in the recent seminal work of de Rham {\it et.al.}\cite{deRham:2010kj} while the latter boils down to a noncommutativity of limits. For this, there are two possibilities:
\begin{itemize}
    \item turning off interactions does not necessarily commute with the massless limit. In other words, in order to resolve the vDVZ discontinuity, it is necessary to go beyond the linearised Fierz-Pauli action, leading to the famed Vainshtein screening mechanism of \cite{VAINSHTEIN1972393}, or 
    \item the massless limit does not commute with the limit of vanishing cosmological constant.
\end{itemize}
Either case will break Birkhoff's theorem resulting in a vDVZ-like discontinuity. To clarify the situation surrounding this discontinuity, in \cite{Moynihan:2017tva} we attempted to re-phrase the discontinuity in the language of scattering amplitudes, largely because these modern on-shell methods are unencumbered by much of the baggage of the usual Lagrangian formulation of the problem. Recently, there have been several advances in our understanding of gravity by harnessing the on-shell scattering amplitudes paradigm \cite{Cachazo:2017jef,Guevara:2017csg,Arkani-Hamed:2017jhn,Guevara:2018wpp,Chung:2018kqs,Carballo-Rubio:2018bmu,Caron-Huot:2018ape,Guevara:2019fsj,Emond:2019crr,Arkani-Hamed:2019ymq,Moynihan:2019bor,Huang:2019cja,Chung:2019yfs,Cristofoli:2019neg,Chung:2019duq,Bjerrum-Bohr:2019kec,Burger:2019wkq,Moynihan:2020gxj}, including approaches based on the double copy \cite{Bern:2010ue,Bern:1998ug,Green:1982sw,Bern:1997nh,Carrasco:2011mn,Carrasco:2012ca,Mafra:2012kh,Boels:2013bi,Bjerrum-Bohr:2013iza,Bern:2013yya,Bern:2013qca,Nohle:2013bfa,
	Bern:2013uka,Naculich:2013xa,Mafra:2014gja,Bern:2014sna,
	Mafra:2015mja,He:2015wgf,Bern:2015ooa,
	Mogull:2015adi,Chiodaroli:2015rdg,Bern:2017ucb,Johansson:2015oia,Oxburgh:2012zr,Geyer:2019hnn,White:2011yy,Melville:2013qca,Luna:2016idw,Saotome:2012vy,Vera:2012ds,Johansson:2013nsa,Johansson:2013aca,Bargheer:2012gv,Huang:2012wr,Chen:2013fya,Chiodaroli:2013upa,Johansson:2014zca,Johansson:2017srf,Cheung:2017ems,Chiodaroli:2017ehv,Chen:2019ywi,Plefka:2019wyg,Aoude:2019xuz,Lipstein:2019mpu,Adamo:2020qru,Borsten:2020xbt} (see \cite{Bern:2019prr} for a recent review), the \textit{classical} double copy \cite{Monteiro:2014cda,Luna:2015paa,Luna:2016due,Goldberger:2016iau,Anastasiou:2014qba,Borsten:2015pla,Anastasiou:2016csv,Anastasiou:2017nsz,Cardoso:2016ngt,Borsten:2017jpt,Anastasiou:2017taf,Anastasiou:2018rdx,LopesCardoso:2018xes,Goldberger:2017frp,Goldberger:2017vcg,Goldberger:2017ogt,Luna:2016hge,Luna:2017dtq,Shen:2018ebu,Levi:2018nxp,Plefka:2018dpa,Cheung:2018wkq,Carrillo-Gonzalez:2018pjk,Monteiro:2018xev,Plefka:2019hmz,Maybee:2019jus,Johansson:2019dnu,PV:2019uuv,Carrillo-Gonzalez:2019aao,Bautista:2019evw,Moynihan:2019bor,Bah:2019sda,CarrilloGonzalez:2019gof,Goldberger:2019xef,Kim:2019jwm,Banerjee:2019saj,Luna:2020adi,Alfonsi:2020lub,Bahjat-Abbas:2020cyb} and those probing the on-shell structure of massive gravity \cite{Momeni:2020vvr,Johnson:2020pny}. In \cite{Moynihan:2017tva}, we utilised this modern approach to show how the vDVZ discontinuity manifests at the level of the scattering amplitudes but did not go quite far enough in disambiguating between the two sources above, mostly because computing massive on-shell amplitudes in gravity at higher (than linear) order was beyond the scope of that article. At the same time we were reminded that this massive to massless discontinuity of the spin-2 graviton is also shared by a spin-$\frac{3}{2}$ Rarita-Schwinger field coupled to a conserved vector-spinor current $j^{\mu}$. This is not an unexpected result since, when the spin-$\frac{3}{2}$ field is coupled to the current, the supermatter interactions resulting from single-fermion interactions have precisely the form required by supersymmetry to complement single-graviton exchange between stress tensor sources \cite{Deser:2000de}. Demonstrating this result using standard methods is a nontrivial exercise in supergravity manipulations \cite{Deser:1977ur}. Furthermore, we are motivated by the fact that, while all evidence so far suggest a massless graviton, the same is not true for the Rarita-Schwinger field. First, we would expect to have observed the gravitino were it massless. Secondly, various supersymmetry breaking mechanisms (such as the super-Higgs effect, or gravitationally induced SUSY breaking) have been shown to endow the Rarita-Schwinger particle with a non-zero mass \cite{Deser:1977uq,Cremmer:1978iv,Cremmer:1982vy}.\\

\noindent
In this article, we approach the spin-$\frac{3}{2}$ discontinuity from the point of view of on-shell massive amplitudes following the analysis in \cite{Moynihan:2017tva}. There it was shown that such a discontinuity can be observed in a purely on-shell manner by directly constructing the massive scattering amplitudes of the theory, taking their massless limit and comparing them with the constructed massless amplitudes. To be concrete, we consider scattering amplitudes in $\cl{N} = 1$ 4D supergravity, whose gauge multiplet consists entirely of a (spin-2) graviton and one Majorana (spin-$\frac{3}{2}$) spinor gravitino. This can be coupled to matter multiplets that preserve the supersymmetry, specifically an $\cl{N} = 1$ vector $(1,\frac{1}{2})$ multiplet consisting of a gauge-boson and gaugino, and an $\cl{N} = 1$ chiral  $(\frac{1}{2},0)$ multiplet consisting of a spin-$\frac{1}{2}$ fermion and a complex scalar.\\
 
\noindent
Giving the graviton a non-zero mass, increases the (on-shell) degrees of freedom from two to five, introducing two vectors and one scalar. On the other hand, giving the gravitino mass results in two additional fermionic modes, corresponding to four on-shell degrees of freedom that are grouped by spin as $\{\pm \frac{3}{2}, \pm \frac{1}{2}\}$. In the case of the massive spin-2 action, the vector modes completely decouple, while the scalar modes couple to the trace of the matter stress energy tensor. In the massless limit then, any coupled matter that has a trace-free stress energy sector won't suffer a discontinuity, while any matter stress energy tensor with a non-zero trace will suffer one. From the on-shell scattering amplitude perspective, this same phenomena can be observed by computing the scattering of scalars via a massive graviton and comparing the massless limit of this amplitude with the same amplitude constructed from a massless graviton. The supersymmetric analog of this statement is that the fermionic modes of the gravitino couple to the Dirac gamma-trace of the current $\gamma_\mu j^\mu$ so in this case, we compare to amplitudes with non-zero gamma-trace.

\section{The Supersymmetric vDVZ Discontinuity}\label{sec:FT}

To begin, we should first clarify what is the analogue of the vDVZ discontinuity in the supersymmetric context. Towards this end, we will utilise the St\"uckelberg formalism, appropriately adapted. Consider then, the free massive Rarita-Schwinger action
\begin{align}\label{key}
S &= -\int d^4x~e\left(\frac12\overline{\Psi}_\mu^\alpha\gamma^{\mu\rho\nu}\pd_\rho\Psi_{\nu\alpha} - \frac{m}{2}\overline{\Psi}_\mu^\alpha\gamma^{\mu\nu}\Psi_{\nu\alpha} + \overline{\Psi}_\mu^\alpha j^\mu_\alpha\right) \equiv -\int d^4x~e~\cl{L},
\end{align}
where $\alpha$ is a spinor index, $\mu,\nu,\rho$ are Lorentz indices and $e$ is, as usual, the determinant of the frame field $e^{a}_{\mu}(x)$. Our $\gamma$-conventions read
\begin{equation}\label{key}
\gamma^{\mu\rho\nu} \equiv i\epsilon^{\mu\sigma\rho\nu}\gamma_5\gamma_\sigma, ~~~~~\gamma^{\mu\nu} \equiv \frac{i}{2}[\gamma^\mu,\gamma^\nu].
\end{equation}

Without the mass term, this action is invariant under
\begin{equation}\label{key}
\Psi^{\alpha}_\mu \longrightarrow \Psi^{\alpha}_\mu + \pd_\mu\chi^\alpha,~~~~~\overline{\Psi}^{\alpha}_\mu \longrightarrow \overline{\Psi}^{\alpha}_\mu + \overline{\chi}^\alpha\overset{\leftarrow}{\pd}_\mu,
\end{equation}
for some spinor $\chi^\alpha$. This symmetry is broken by the mass term but can be restored if we introduce the supercovariant derivative
\begin{equation}
    D_\mu = \partial_\mu + \frac12 m\gamma_\mu,
\end{equation}
in terms of which the Lagrangian becomes
\begin{equation}
    \cl{L} = \frac12\overline{\Psi}_\mu^\alpha\gamma^{\mu\rho\nu}D_\rho\Psi_{\nu\alpha} + \overline{\Psi}_\mu^\alpha j^\mu_\alpha,
\end{equation}
after judicious use of the identity $\epsilon^{\mu\nu\rho\sigma}\gamma_5\gamma_\nu = 2\gamma^\mu\gamma^{\rho\sigma}$.
We can now introduce $\chi$ as a (spinorial) St\"uckelberg field via the transformation\footnote{The factor of $\frac{1}{\sqrt{3}m}$  ensures a canonical fermionic kinetic term.}
\begin{equation}
    \Psi_\mu^\alpha \longrightarrow \Psi_\mu^\alpha + \frac{1}{\sqrt{3}m}D_\mu\chi^\alpha.
\end{equation}
Under this transformation, the Lagrangian density varies as
\begin{equation}
    \delta\cl{L} = -\frac{\sqrt{3}m}{2}\left(\overline{\slashed{\Psi}}^\alpha\chi_\alpha + \overline{\chi}^\alpha\slashed{\Psi}_\alpha\right) - \overline{\chi}^\alpha(\slashed{\pd} + m)\chi_\alpha + \frac{1}{\sqrt{6}}\overline{\chi}^\alpha \slashed{j}_\alpha\,.
\end{equation}
Subsequently, taking the massless limit does not lead to the original massless Lagrangian since,
\begin{equation}
    \cl{L}^{massive}\bigg|_{m\longrightarrow 0} = \cl{L}^{massless} + \frac{1}{\sqrt{6}}\overline{\chi}^\alpha \slashed{j}_\alpha.
\end{equation}
It is this that we identify as the SUSY equivalent of the vDVZ discontinuity, since any matter with a $\gamma$-traceless current will couple differently than matter with a non-vanishing $\gamma$-trace.

As alluded to in the introduction, we wish to couple to two chiral multiplets: a scalar multiplet ($\frac{1}{2},0$) and a vector multiplet $(1,\frac{1}{2})$. The corresponding vector-spinor currents are, by Noethers theorem, 
\begin{align}
j_{\alpha}^\mu[\Phi,\psi] &= \left[i\gamma^\nu\pd_\nu (\phi_1 -i\gamma_5\phi_2) -m(\phi_1 + i\gamma_5\phi_2)\right]\gamma^\mu\psi_\alpha\\
    j_{\alpha}^\mu[A,\lambda] &= \gamma^\rho\gamma^\nu\gamma^\mu \lambda_\alpha F_{\rho\nu},
\end{align}

where $\Phi = \phi_1 + i\phi_2$ is a complex scalar, $\psi_\alpha$ a Majorana fermion (both with mass $m_\Phi$), $\lambda_\alpha$ a massless photino and $F_{\rho\nu}$ the Maxwell tensor for photon $A_\nu$. These are both conserved, i.e. that $\pd_\mu j_{\alpha}^\mu[\Phi,\psi] = \pd_\mu j_{\alpha}^\mu[A,\lambda] = 0$, however only one has a non-zero Dirac trace, e.g.
\begin{equation}
    \slashed{j}_{\alpha}[\Phi,\psi] \neq 0,~~~~~\slashed{j}_{\alpha}[A,\lambda] = 0.
\end{equation}
While this formulation is clearly off-shell, it will inform the on-shell investigation to which we now turn.  
\section{$\cl{N} = 1$ Supersymmetric Discontinuity} \label{sec:susy}
Let's start by thinking about the supermultiplets. To ensure that we preserve the correct symmetries we have to let all the particles in the multiplet have the same mass. To this end we will draw on the vector- and scalar-multiplet currents as they are stated in the previous section. In the interest of self-containment, many of the techniques and conventions used throughout this chapter will be set up in this first section.

\subsection{Bold Notation and the St\"uckelberg Decomposition}
As described in \cite{Arkani-Hamed:2017jhn}, we can significantly reduce the notational overhead that comes with incorporating $SU(2)$ little group indices in favour of bold-facing the massive particle spinors. Given our treatment of the action in section \ref{sec:FT}, it will be useful to perform a similar analysis at the level of the on-shell three-particle amplitudes. To this end, consider the coupling of a massless photon and massless fermion to a massive gravitino. The associated amplitude is given by
\begin{equation}
   \cl{M}_3^{\{IJK\}}[1^{-1/2},2^{-1},\textbf{3}^{3/2}] = \frac{\kappa}{m}\braket{1\textbf{3}}\braket{2\textbf{3}}^2.
\end{equation}
Simply unbolding this expression does {\it not} yield the correct result, since taking the massless limit tells us that the spin $-3/2$ mode of the gravitino diverges. Instead, we need to write it in a form where we \textit{can} unbold, i.e.
\begin{equation}
\begin{split}
\cl{M}_3^{\{IJK\}}[1^{-1/2},2^{-1},\textbf{3}^{3/2}]
&=
\frac{\kappa}{m^2} \bra{1}p_2\sket{\bt{3}}\ang{2\bt{3}}^2\\
&=
-\kappa \frac{\ang{12}\ang{2\bt{3}}^2\squ{\bt{3}2}}{\ang{2\bt{3}^{I}}\squ{\bt{3}_{I}2}}.
\end{split}
\end{equation}
Now, when we unbold, we find that the longitudinal spin $1/2$ mode of the gravitino is non-vanishing in the massless limit, since
\begin{equation}\label{key}
\cl{M}_3^{\{IJK\}}[1^{-1/2},2^{-1},\textbf{3}^{3/2}]\bigg|_{m\to 0} = -\frac{\kappa}{3}\braket{12}\braket{23},
\end{equation}
and with the inclusion of the relevant symmetry factor. This is consistent with the field theory treatment. On the other hand, we find that the $3/2$ mode does indeed survive, but for a different helicity choice of the fermion and photon
\begin{equation}\label{key}
\cl{M}_3^{\{IJK\}}[1^{+1/2},2^{-1},\bt{3}^{3/2}] = \kappa \frac{\ang{2\bt{3}}^3}{\ang{12}} ~~\longrightarrow~~ \kappa \frac{\ang{23}^3}{\ang{12}}.
\end{equation}
This implies that the act of expanding three-paticle amplitudes into their various helicity components is the on-shell avatar of the St\"uckelberg decomposition, reflecting our result in section \ref{sec:FT}. This example manifests a subtlety in this approach; for the `unbolding' of massive amplitudes to be physically meaningful, we will often have to make sure that we have teased out as much of the explicit mass dependence as possible, a point that will be important for what follows. Another, perhaps more easily digested, example is the three-particle amplitude of a massive vector and two massless scalars. This amplitude is treated in detail in \cite{Arkani-Hamed:2017jhn} by making appropriate choices for the massive spinor indices to reveal the underlying helicities in the massless limit. The amplitude can be written in the generic form 
\begin{equation}
	\cl{A}^{\{IJ\}}[1^0,2^0,\bt{3}^1]\propto \frac{\bra{\bt{3}}p_1p_2\ket{\bt{3}}}{m^2} + \frac{\bra{\bt{3}}p_1\sket{\bt{3}}}{m} + \frac{\sbra{\bt{3}}p_1p_2\sket{\bt{3}}}{m^2}.
\end{equation}
Naively unbolding any of these terms to take the massless limit appears to be divergent. However, using the fact that $m^2 = \ang{12}\squ{12}$ it can be rewritten as
\begin{equation}
	\cl{A}^{\{IJ\}}[1^0,2^0,\bt{3}^1]\propto \frac{\ang{\bt{3}1}\ang{2\bt{3}}}{\ang{12}} + m\frac{\ang{\bt{3}1}\squ{1\bt{3}}}{\ang{1\bt{3}^I}\squ{\bt{3}_I 1}} + \frac{\squ{\bt{3}1}\squ{2\bt{3}}}{\squ{12}}.
\end{equation}
Now that the mass factors in the denominators have been taken care of we can simply unbold and retain the term with the appropriate helicity. The first term is clearly the $h_3=-1$ helicity mode, the last term the $h_3=+1$ helicity mode and the scalar mode vanishes in the massless limit as expected. We see that we can simply extract the appropriate modes of the vector in the massless limit by finding the explicit mass dependence of a general form of the amplitude and then simply unbolding.
\subsection{Vector Mulitplet} \label{sec:susyvector}

\begin{figure}[h]
	\centering
	\begin{tikzpicture}[scale=1]
	\begin{feynman}  
	\vertex (a) at (-4,2) {$p_1^{\pm 1/2}$};
	\vertex (b) at (-4,-2) {$p_2^{-1}$};
	\vertex (c) at (2,-2) {$p_3^{\mp 1/2}$};
	\vertex (d) at (2,2) {$p_4^{+1}$};
	\vertex (r) at (0,0);
	\vertex (l) at (-2,0) ;
	\diagram* {
		(a) -- [fermion] (l) -- [graviton] (r) -- [photon] (d),
		(b) -- [photon] (l) -- [graviton] (r) -- [fermion] (c),
		(l)--[plain] (r)
	};
	\draw (-1,-0.1) node[below] {$\bt{p}^{3/2}$};
	\end{feynman}
	\end{tikzpicture}
	\caption{Vector multiplet four-particle}
	\label{diag:vectorSUSY}
\end{figure}
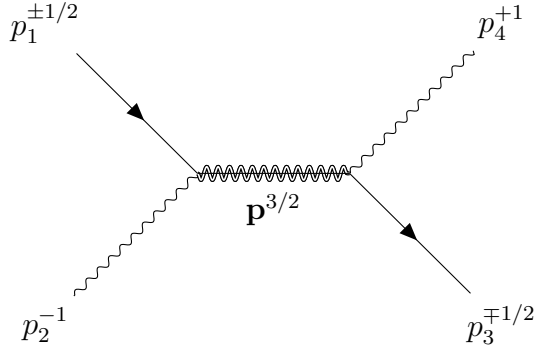

We begin by computing the amplitudes involving an $\cl{N} = 1$ massless vector multiplet  $(1,\frac{1}{2})$. This produces a three-particle vertex consisting of a gauge boson (a photon), a fermion (the photino) and a gravitino interacting with a coupling $\kappa $. We'll require  the propagating gravitino to be massive in order to study the effects of taking the massless limit. The associated four-particle diagram is given by Fig. \ref{diag:vectorSUSY}.

We need only construct the left hand three-particle amplitude for all the possible helicity configurations. This is because the right hand three-particle amplitudes can be obtained by complex conjugation of eqs. \eqref{eq:susyvector3point1} and \eqref{eq:susyvector3point2}, and making the replacements $p_1\to p_3, ~p_2\to p_4$. Throughout this article we will use the formalism developed in Ref. \cite{Arkani-Hamed:2017jhn}, including \textbf{bold} notation to highlight when a spinor represents a massive particle, suppressing the $SU(2)$ little group indices unless required for clarity. Using this formalism, we find that the possible three-particle amplitudes are 

\begin{align}
\cl{M}_3^{\lbrace JKL \rbrace}[1^{+1/2},2^{-1},\bt{p}^{3/2}]
&=
\frac{\kappa}{m^2} \ang{2\bt{p}}^3 \squ{12}
=
\kappa \frac{\ang{2\bt{p}}^3}{\ang{12}},\label{eq:susyvector3point1}\\
\cl{M}_3^{\lbrace JKL \rbrace}[1^{-1/2},2^{-1},\bt{p}^{3/2}]
&=
\frac{\kappa}{m} \ang{1\bt{p}}\ang{2\bt{p}}^2.\label{eq:susyvector3point2}
\end{align}
From this it is now simple to compute the two possible four-particle amplitudes by contracting the massive indices of the internal particle 
\begin{equation}\label{eq:susy4vector1}
\begin{split}
\cl{M}_4[1^{+1/2},2^{-1},3^{-1/2},4^{+1}]
&=\cl{M}^{\lbrace IJK\rbrace}[1^{+1/2},2^{-1},\bt{p}^{3/2}]\frac{1}{p^2+m^2} \tilde{\cl{M}}_{\lbrace IJK\rbrace}[3^{-1/2},4^{+1},-\bt{p}^{3/2}]\\
&= -\frac{\kappa^2}{t+m^2} \frac{\squ{14}}{\squ{34}}\bra{2}p_1\sket{4}^2 \\&= \frac{\squ{14}}{\squ{34}}\cl{M}_4[1^{0},2^{-1},3^0,4^{+1}],
\end{split}
\end{equation}
where $\cl{M}_4[1^{0},2^{-1},3^0,4^{+1}]$ is the massive graviton mediated scalar-photon amplitude \cite{Moynihan:2017tva}. This explicitly shows that the supersymmetric Ward identity is satisfied, as expected. This amplitude only has an explicit mass dependence in the propagator, and we can therefore easily take the massless limit. We want to compare the massless limit of the amplitude with the four-particle amplitude with an initially massless gravitino, which must be constructed from the three-particle amplitude
\begin{equation}
\label{eq:vectormassless}
\cl{M}_3[1^{+1/2},2^{-1},p^{-3/2}] = \kappa \frac{\ang{2p}^3}{\ang{12}}.
\end{equation}
Using this, we find that the massless four-particle amplitude is
\begin{align}\label{eq:susy4vectormassless}
\cl{M}_4[1^{+1/2},2^{-1},3^{-1/2},4^{+1}]\bigg|_{m=0}
&= \sum_{\pm} \cl{M}_3[1^{+1/2},2^{-1},p^{\pm 3/2}] \frac{1}{p^2} \cl{M}_3[3^{-1/2},4^{+1},-p^{\mp 3/2}]\nonumber\\
&=-\frac{\kappa^2}{t} \frac{\squ{14}}{\squ{34}}\bra{2}p_1\sket{4}^2,
\end{align}
 in agreement with eq. \eqref{eq:susy4vector1}. Explicitly,
\begin{equation}
\cl{M}_4[1^{+1/2},2^{-1},3^{-1/2},4^{+1}]\bigg|_{m\to 0} = \cl{M}_4[1^{+1/2},2^{-1},3^{-1/2},4^{+1}]\bigg|_{m=0}.
\end{equation}
If our only concern was reproducing the field-theory result in section \ref{sec:FT} we could well stop at this point, noting that this is the only \textit{chiral} amplitude possible for this field configuration. However, we do not wish to be led by the field theory construction, and so with no {\it a priori} reason to discard the other helicity possibility, we have include it. To construct the non-chiral four-particle amplitude,  notice that the three-particle amplitude in eq.\eqref{eq:susyvector3point2} is symmetric in two of its indices and, summing over the internal $SU(2)$ indices, allows us to write 
\begin{eqnarray*}
\cl{M}_3^{\{JKL\}}\cl{M}_{3\{JKL\}} = \frac{1}{6}\cl{M}_3^{JKL}\cl{M}_{3\{JKL\}} = \frac{1}{3}\cl{M}_3^{JKL}(\cl{M}_{3JKL}+\cl{M}_{3KJL}+\cl{M}_{3LKJ}),
\end{eqnarray*}
which in turn permits the four-particle amplitude to be written as
\begin{equation}\label{eq:susy4vector2}
\begin{split}
\cl{M}_4[1^{-1/2},2^{-1},3^{+1/2},4^{+1}]
&=\cl{M}_3^{\lbrace IJK\rbrace}[1^{-1/2},2^{-1},\bt{p}^{3/2}]\frac{1}{p^2+m^2} \tilde{\cl{M}}_{3\lbrace IJK\rbrace}(3^{+1/2},4^{+1},-\bt{p}^{3/2})\\
&= \frac{\kappa^2}{3 m^2(p^2+m^2)}( \bra{1}p\sket{3}\bra{2}p\sket{4}^2+2\bra{1}p\sket{4}\bra{2}p\sket{3}\bra{2}p\sket{4})\\
&= -\frac{\kappa^2}{3(p^2+m^2)}\ang{12}\squ{34}\bra{2}p\sket{4} + \cl{O}(m^2)\,,
\end{split}
\end{equation}
where, in the last line we have used the Schouten identity\footnote{To see that the second piece of the amplitude is $\propto m^4$, multiply by $\squ{12}^2/\squ{12}^2$ and use conservation of momentum.}. Now we can easily take the massless limit to find
\begin{equation}
\begin{split}
\cl{M}_4[1^{-1/2},2^{-1},3^{+1/2},4^{+1}]\bigg|_{m\to0}
&=
-\frac{\kappa^2}{3t} \ang{12}\squ{34}\bra{2}p\sket{4} \\&= \frac13\cl{M}_4[1^{-1/2},2^{-1},3^{+1/2},4^{+1}]\bigg|_{m=0}
\end{split}
\end{equation}
and where 
\begin{eqnarray*}
\cl{M}_4[1^{-1/2},2^{-1},3^{+1/2},4^{+1}]\bigg|_{m=0} = \cl{M}_3[1^{-1/2},2^{-1},p^{+1/2}]\cl{M}_3[3^{+1/2},4^{+1},-p^{-1/2}]/t\,.
\end{eqnarray*}
From this it is clear that this helicity structure produces a contribution due to the spin-$\frac{1}{2}$ mode of the gravitino but with an overall factor of a $1/3$ when compared with its massless counterpart.

\subsection{Scalar Mulitplet} \label{sec:susyscalar}

\begin{figure}[h]
	\centering
	\begin{tikzpicture}[scale=1]
	\begin{feynman}  
	\vertex (a) at (-4,2) {$\bt{p}_1^{1/2}$};
	\vertex (b) at (-4,-2) {$\bt{p}_2^{0}$};
	\vertex (c) at (2,-2) {$\bt{p}_3^{1/2}$};
	\vertex (d) at (2,2) {$\bt{p}_4^{0}$};
	\vertex (r) at (0,0);
	\vertex (l) at (-2,0) ;
	\diagram* {
		(a) -- [fermion] (l) -- [graviton] (r) -- [plain] (d),
		(b) -- [plain] (l) -- [graviton] (r) -- [fermion] (c),
		(l)--[plain] (r),
	};
	\draw (-1,-0.1) node[below] {$\bt{p}^{3/2}$};
	\end{feynman}
	\end{tikzpicture}
	\caption{Scalar multiplet four-particle}
	\label{diag:scalarSUSY}
\end{figure}
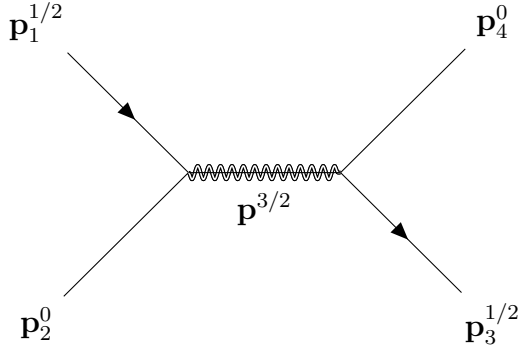

For the scalar multiplet, we need to consider a massive gravitino mediated interaction of a massive fermion and a massive scalar. For simplicity, we will take the masses of the latter two to be the same. The corresponding four-particle diagram is given in Fig. \ref{diag:scalarSUSY}.

Following Ref. \cite{Arkani-Hamed:2017jhn} the all-massive three-particle amplitude can be constructed as
\begin{equation}
\begin{split}
\cl{M}_3^{I\lbrace JKL\rbrace}[\bt{1}^{1/2},\bt{2}^0,\bt{p}^{3/2}]
&=
g_1 \ang{\bt{1}\bt{p}}\ang{\bt{p}\bt{p}} \\
&+g_2 (\ang{\bt{1}\bt{p}}\bra{\bt{p}}p_1p\ket{\bt{p}}
+\ang{\bt{p}\bt{p}}(\bra{\bt{1}}p_1p\ket{\bt{p}}
+\bra{\bt{p}}p_1p\ket{\bt{1}}))\\
&+g_3 (\bra{\bt{1}}p_1p\ket{\bt{p}}
+\bra{\bt{p}}p_1p\ket{\bt{1}})\bra{\bt{p}}p_1p\ket{\bt{p}}),
\end{split}
\end{equation}
where we have chosen to expand in a basis $(\mathcal{O}^{ab} = p_1^{\lbrace a\db}p_{\db}^{b\rbrace}, \epsilon^{ab})$ and the coupling functions\footnote{The `coupling functions' are functions of particle mass and the gravitational coupling $\kappa$ only, with the mass dependence needing to be fixed by e.g. demanding the correct high-energy behaviour} $g_i$ have an undetermined mass dependence that will be fixed shortly. To simplify this, note that since the external $SU(2)$ indices of the gravitino have to be symmetrised, any term that has a factor $\ang{\bt{p}^I\bt{p}^J}=m\epsilon^{IJ}$ (where $IJ$ are free indices) will vanish once symmetrised and can be ignored. Taking the external particle masses to be $m_1$ and the internal gravitino to have mass $m$, reduces the above amplitude to
\begin{equation} \label{eq:susyscalar3massive}
\begin{split}
\cl{M}_3^{I\lbrace JKL\rbrace}(\bt{1}^{1/2},\bt{2}^0,\bt{p}^{3/2})
&=
m(-g_2+g_3m^2)\ang{\bt{1}\bt{p}}\bra{\bt{p}}p_1\sket{\bt{p}}
+2m^2m_1g_3\squ{\bt{1}\bt{p}}\bra{\bt{p}}p_1\sket{\bt{p}}.
\end{split}
\end{equation}
In getting to this point we have made liberal use of the identities in appendix \ref{app:ident}. In order to fix the coupling functions $g_i$, we can compare these amplitudes with the ones derived in limit where one particle becomes massless. Taking the external mass to zero then, we find
\begin{equation}
\cl{M}_3^{I\lbrace JKL\rbrace}[\bt{1}^{1/2},\bt{2}^0,\bt{p}^{3/2}]\bigg|_{m_1\to 0}
=
\begin{cases}
\frac{\kappa}{m} \ang{1\bt{p}}\bra{\bt{p}}p_1\sket{\bt{p}} ,~~~h_1=-1/2\\
\frac{\kappa}{m} \squ{1\bt{p}}\bra{\bt{p}}p_1\sket{\bt{p}} ,~~~h_1=+1/2,
\end{cases}
\end{equation}
which implies that the following limits must hold 
\begin{equation}\label{eq:3ptlims}
\lim_{m_1 \to 0} m(-g_2+g_3m^2) = \frac{\kappa}{m},~~~~~\lim_{m_1\to 0}2m^2m_1g_3 = \frac{\kappa}{m}.
\end{equation}
Next, to isolate any $m_1$ dependence, take $m\to0$ so that
\begin{equation}
\cl{M}_3^{I\lbrace JKL\rbrace}[\bt{1}^{1/2},\bt{2}^0,\bt{p}^{3/2}]\bigg\vert_{m\to 0}
\simeq
\begin{cases}
\kappa m_1 \ang{\bt{1}p} ,~~~h_p=-1/2,\\
\kappa \frac{\ang{\bt{1}p}\bra{p}p_1\sket{\xi}}{\squ{p\xi}} ,~~~h_p=-3/2,\\
\kappa m_1 \squ{\bt{1}p} ,~~~h_p=+1/2,\\
\kappa \frac{\squ{\bt{1}p}\bra{\xi}p_1\sket{p}}{\ang{\xi p}} ,~~~h_p=+3/2
\end{cases}
\end{equation}
Now we can, for example, look at the $-3/2$ mode of the gravitino, choose $\xi = \eta$ and compare with the $I=J=K=1$ amplitude,
\begin{eqnarray}
\cl{M}_3^{I\lbrace 111\rbrace}[\bt{1}^{1/2},\bt{2}^0,\bt{p}^{3/2}] = \kappa \frac{\ang{\bt{1}p}\bra{p}p_1\sket{\eta}}{m} + \cl{O}(m).
\end{eqnarray}
Recognising that in the massless limit $m \longrightarrow [p\eta]$, implies that the limits in eq. \eqref{eq:3ptlims} must hold. In other words,  the amplitude has no explicit $m_1$-dependence (although it will have {\it implicit} dependence on $m_{1}$ to recover e.g. the $h_p = \pm 1/2$ amplitudes). Putting this together then, the three-particle amplitude must take the form
\begin{equation} \label{eq:susyscalar3massive1}
\begin{split}
	\cl{M}_3^{I\lbrace JKL\rbrace}\left[\bt{1}^{1/2},\bt{2}^0,\bt{p}^{3/2}\right]
	&=
	\frac{\kappa}{m}(\ang{\bt{1}\bt{p}}+\squ{\bt{1}\bt{p}})\bra{\bt{p}}p_1\sket{\bt{p}}.
\end{split}
\end{equation}
With the three-particle amplitude in hand, we can now construct the four-particle amplitude. Given its lack of index symmetry however, the four-particle amplitude will have six distinct tensor structures and is particularly unwieldy. We will spare the reader the gory details. Suffice it to say that, with judicious use of the identities in appendix \ref{app:ident}, it can be written as
\begin{eqnarray}\label{eq:susyscalar4massive2}
\cl{M}_4^{I_1I_3}\left[\bt{1}^{1/2},\bt{2}^{0},\bt{3}^{1/2},\bt{4}^{0}\right]
&=&
\frac{\kappa^2}{6\bt{t}m^2}
(\bra{\bt{1}}p\sket{\bt{3}} - \bra{\bt{3}}p\sket{\bt{1}})(-6m^2(2p_1\cdot p_3) +2 (2p\cdot p_1)^2\nonumber\\
&+&4m^2m_1^2) -4mm_1^2(\bra{\bt{1}}p\sket{\bt{3}} + \bra{\bt{3}}p\sket{\bt{1}})(2p\cdot p_1)\\
&+&4m(\ang{\bt{13}} + \squ{\bt{13}})(2p\cdot p_1)^2) +\mathcal{O}(m).\nonumber
\end{eqnarray}
Now, using the fact that $2p\cdot p_1=-2p\cdot p_3= -m^2$, we take the massless limit to get,
\begin{equation}\label{eq:susyscalar4massive-m0}
\begin{split}
\cl{M}_4^{I_1I_3}\left[\bt{1}^{1/2},\bt{2}^{0},\bt{3}^{1/2},\bt{4}^{0}\right]\bigg\vert_{m\to0}
&=
-\frac{\kappa^2}{t}(\bra{\bt{1}}p\sket{\bt{3}}-\bra{\bt{3}}p\sket{\bt{1}})(2p_1 \cdot p_3 -\frac{2}{3} (m_1^2))\,.
\end{split}
\end{equation}
This in turn needs to be compared to the four-particle amplitude with a massless gravitino exchange, 
\begin{eqnarray}\label{eq:susyscalar4massless}
\cl{M}^{3/2}_4{}^{I_1I_3}[\bt{1}^{1/2},\bt{2}^{0},\bt{3}^{1/2},\bt{4}^{0}]\bigg\vert_{m=0}
\!\!\!\!&=&
\sum_{\pm3/2}\cl{M}_3^{I_1\lbrace JKL\rbrace}(\bt{1}^{1/2},\bt{2}^0,p^{\pm3/2})\frac{1}{t}\cl{M}_3^{I_3}{}_{\lbrace JKL\rbrace}(\bt{3}^{1/2},\bt{4}^0,-p^{\mp3/2})\nonumber\\
&=&
\frac{\kappa^2}{t}(\bra{\bt{1}}p\sket{\bt{3}}\frac{\bra{\zeta}p_3 p p_1\sket{\xi}}{\bra{\zeta}p\sket{\xi}}-\mathrm{C.C.}),
\end{eqnarray}
where $\zeta$ and $\xi$ are reference spinors. To facilitate this comparison, we choose $\zeta=\xi$ and utilise conservation of momentum and the Schouten identity to write the amplitude as
\begin{eqnarray}\label{eq:susyscalar4massless2}
\cl{M}^{3/2}_4{}^{I_1I_3}[\bt{1}^{1/2},\bt{2}^{0},\bt{3}^{1/2},\bt{4}^{0}]\bigg\vert_{m=0}
&=&
-\frac{\kappa^2}{t}(\bra{\bt{1}}p\sket{\bt{3}}-\bra{\bt{3}}p\sket{\bt{1}})(2p_1\cdot p_3)\,.
\end{eqnarray}
To make a concrete comparison of \eqref{eq:susyscalar4massive-m0} with its massless propagator counterparts we also need to compute the four particle amplitude with a massless fermion exchange. Using the three particle amplitudes given in \cite{Arkani-Hamed:2017jhn} this is straightforwardly computed as 
\begin{eqnarray}\label{eq:susyscalar4massless-12}
\cl{M}^{1/2}_4{}^{I_1I_3}[\bt{1}^{1/2},\bt{2}^{0},\bt{3}^{1/2},\bt{4}^{0}]\bigg\vert_{m=0}
&=&
\sum_{\pm1/2}\cl{M}_3^{I_1\lbrace JKL\rbrace}(\bt{1}^{1/2},\bt{2}^0,p^{\pm1/2})\frac{1}{t}\cl{M}_3^{I_3}{}_{\lbrace JKL\rbrace}(\bt{3}^{1/2},\bt{4}^0,-p^{\mp1/2})\nonumber\\
&=&
\frac{\kappa^2}{t}(\bra{\bt{1}}p\sket{\bt{3}}-\bra{\bt{3}}p\sket{\bt{1}})(m_1^2).
\end{eqnarray}
We are now in a position to write the amplitude \eqref{eq:susyscalar4massive-m0} in terms of its massless counterparts \eqref{eq:susyscalar4massless2} and \eqref{eq:susyscalar4massless-12} as
\begin{equation}
	\cl{M}_4^{I_1I_3}[\bt{1}^{1/2},\bt{2}^{0},\bt{3}^{1/2},\bt{4}^{0}]\bigg\vert_{m\to0} \!\!\!\!\!\!= \cl{M}^{3/2}_4{}^{I_1I_3}[\bt{1}^{1/2},\bt{2}^{0},\bt{3}^{1/2},\bt{4}^{0}]\bigg\vert_{m=0} + \frac{2}{3}\cl{M}^{1/2}_4{}^{I_1I_3}[\bt{1}^{1/2},\bt{2}^{0},\bt{3}^{1/2},\bt{4}^{0}]\bigg\vert_{m=0}.
\end{equation}
From this expression, it is evident that a discontinuity should be expected in the scalar sector, in precisely the same form as from the field theory analysis. Furthermore, taking the external particle masses to zero and making a specific helicity choice shows that the four-particle amplitude satisfies the expected Ward identity
\begin{equation}\label{key}
\cl{M}_4[1^{-1/2},2^{0},3^{+1/2},4^{0}] = \frac{\braket{12}}{\braket{32}}\cl{M}_4[1^{0},2^{0},3^{0},4^{0}],
\end{equation}
where $\cl{M}_4[1^{0},2^{0},3^{0},4^{0}]$ is the massless limit of eq. 3.18 in Ref \cite{Moynihan:2017tva}.

\section{Supersymmetry Breaking and the Discontinuity}
An interesting question that arises from our analysis above is whether this discontinuity persists below the supersymmetry breaking scale. Indeed, from the field theory perspective, none of the arguments for the existence of the discontinuity depend in any substantial way on whether the supersymmetry is broken or not,  just that a massive spin-$\frac{3}{2}$ propagator couples to a current with $j\cdot\gamma = 0$ which is then compared to one for which $j\cdot \gamma \neq 0$. Our on-shell analysis however provides a simple test of this hypothesis by considering multiplets with distinct masses, {\it e.g.} a massive scalar and fermion with masses $m_s \neq m_f$. 

The natural multiplets to consider in this case are,
\begin{enumerate}
   \item the vector multiplet with a massless photon and a fermion of mass $m_f$ and 
   \item the scalar multiplet with a massive scalar and massive fermion with respective masses $m_s \neq m_f$.
 \end{enumerate}  
However there is a subtlety that requires some discussion. In the previous case with unbroken supersymmetry, the vector multiplet amplitudes only contain a single mass with which to constrain the coupling function, and while the scalar multiplet contains two distinct masses, there are well defined massless amplitudes to compare to (after taking appropriate massless limits), thereby allowing us to derive the correct mass structure. Breaking supersymmetry, on the other hand, results in even more masses that could form part of the coupling function. In this case, little group scaling and dimensional analysis alone are insufficient to constrain the masses.

For example, in the scalar multiplet where only the fermion is massive, according to the one-massive formula given in Ref. \cite{Arkani-Hamed:2017jhn}, the amplitude $\cl{M}_3^{I}[\bt{1}^{1/2},2^{0},p^{\pm3/2}] = 0$. This is due to the requirement that $S+h_1-h_2$ and $S+h_2-h_1$ must both be positive, which is clearly not the case if $S = 1/2, h_1 = 0$ and $h_2 = \pm 3/2$. On the other hand, the fully massless amplitude \textit{does} exist and is given by 
\begin{equation}
\cl{M}_3[1^{-1/2},2^{0},p^{-3/2}] = \kappa \frac{\ang{1p}^2\ang{2p}}{\ang{12}}\label{incon1}\,.
\end{equation}
This is obviously not a limit of the one-massive case above, a fact that is also true if only the scalar is left massive. It seems then that the only consistent way to fix the mass dependence of the coupling function $g_i$ is to demand that every amplitude has a well defined massless limit and discarding it if it doesn't. To constrain such coupling functions, let's consider the two-massive-one-massless three-particle amplitude, taking care to retain the gravitino as one of the massive particles. We will compare this to the amplitude with \textit{only} the gravitino massive
\begin{equation}
\cl{M}_3^{\{IJK\}}[1^{-1/2},2^{0},\bt{p}^{3/2}] = \frac{\kappa}{m} \ang{1\bt{p}}\bra{\bt{p}}p_1\sket{\bt{p}}\label{incon5}
\end{equation}
since this does recover the massless amplitude in eq. \eqref{incon1}. Demanding that we recover this amplitude from the two-massive amplitudes
\begin{equation}
	\begin{split}
		\cl{M}_3^{I\lbrace JKL\rbrace}[\bt{1}^{1/2},2^{0},\bt{p}^{3/2}] &= \frac{\kappa}{m} (\ang{\bt{1}\bt{p}}\bra{\bt{p}}p_1\sket{\bt{p}} +\squ{\bt{1}\bt{p}}\bra{\bt{p}}p_1\sket{\bt{p}})\\
		\cl{M}_3^{\lbrace JKL\rbrace}[1^{-1/2},\bt{2}^{0},\bt{p}^{3/2}] &= \frac{\kappa}{m} \ang{1\bt{p}}\bra{\bt{p}}p_1\sket{\bt{p}},
	\end{split}
\end{equation}
constrains the three-massive particle amplitude to the form,
\begin{equation}
	\cl{M}_3^{I\lbrace JKL\rbrace}[\bt{1}^{1/2},\bt{2}^0,\bt{p}^{3/2}]
	=
	\frac{\kappa}{m}(\ang{\bt{1}\bt{p}}+\squ{\bt{1}\bt{p}})\bra{\bt{p}}p_1\sket{\bt{p}}.
\end{equation}
The construction of the four-particle amplitude now mirrors that outlined in the previous section exactly, with the result that
\begin{eqnarray}\label{eq:scalar4massive}
	\cl{M}_4^{I_1I_3}[\bt{1}^{1/2},\bt{2}^{0},\bt{3}^{1/2},\bt{4}^{0}]
	&=&
	\frac{\kappa^2}{6tm^2}
	((\bra{\bt{1}}p\sket{\bt{3}} - \bra{\bt{3}}p\sket{\bt{1}})(-6m^2(2p_1\cdot p_3) +2 (2p\cdot p_1)^2\nonumber\\
	&+&4m^2m_f^2) -4mm_f^2(\bra{\bt{1}}p\sket{\bt{3}} + \bra{\bt{3}}p\sket{\bt{1}})(2p\cdot p_1)\\
	&+&4m(\ang{\bt{13}} + \squ{\bt{13}})(2p\cdot p_1)^2) +\mathcal{O}(m).\nonumber
\end{eqnarray}
The difference between this expression and \eqref{eq:susyscalar4massive-m0} is that $2p\cdot p_1 = m^2 +m_f^2 -m_s^2 = -2p\cdot p_3$. Correspondingly, the amplitude \eqref{eq:scalar4massive} will have terms of order $\cl{O}(m^{-1})$ and therefore has no consistent massless limit. This appears to be an artefact of constructing massive three-particle amplitudes in this formalism, and specifically fixing the mass structure of the arbitrary coupling functions. If this inconsistency were somehow resolved, we  would expect the mass factor in the denominator to have an additive form, $m+m_f-m_s$, which would allow the massless limit to be taken in the four-particle amplitude. That said, it is not clear to us how such a denominator could arise from a local quantum field theory with polynomial interactions. For example, an action containing a term of the form $\sim \phi\slashed{\Psi}^\alpha\psi_\alpha$ will have at most a single mass in the denominator coming from the Rarita-Schwinger polarization. In this sense, a mass dependence of $\cl{O}(m^{-1})$ is to be expected and therefore the formalism does seemingly produce the correct amplitude, but one which apparently doesn't have well defined massless limits. It is entirely plausible that this problem is cured by a high energy mechanism, e.g. the Higgs, which does in some circumstances restore the high energy limit via the goldstone equivalence theorem. The gravitino-goldstino equivalence theorem shows that spontaneously broken symmetries involving longitudinally polarized gravitinos can have a well defined high energy behaviour via the goldstone equivalence \cite{Casalbuoni:1988kv}. However, it is not clear to us that this will solve all of the issues raised in this section, and is at any rate beyond the scope of this article and so we leave this investigation to the future.

Amplitudes for the vector multiplet exhibit a similar inconsistency in that the amplitude with a single massive photino does not reduce to the required massless limit. Specifically,
\begin{equation}\label{key}
\cl{M}_3^{I}[\textbf{1}^{1/2},2^{-1},p^{-3/2}] = \frac{\kappa}{m_f}\braket{\textbf{1}p}\braket{p2}^2
\end{equation}
cannot reproduce the amplitude $\cl{M}_3[1^{-1/2},2^{-1},p^{-3/2}] = 0$ in the limit that $m_f\rightarrow 0$. Again we turn to the case where the gravitino is the only massive particle to fix the coupling function. The corresponding amplitude,
\begin{align} \label{eq:vector2}
M^{I \lbrace JKL \rbrace}[\bt{1}^{ 1/2},2^{-1},\bt{p}^{3/2}]
&=
\frac{\kappa}{m} (\squ{\bt{1}\bt{p}}\ang{\bt{p}2}^2 + \ang{\bt{1}\bt{p}}\ang{\bt{p}2}^2),
\end{align}
correctly reproduces all of the expected three-particle amplitudes in the relevant limits. From this, and with some straightforward but tedious simplification, we find the four-particle amplitude
\begin{eqnarray}\label{eq:vector4massive}
\cl{M}_4^{I_1I_3}[\bt{1}^{1/2},2^{-1},\bt{3}^{1/2},4^{-1}]
&=&
\frac{\kappa^2}{3tm^2}
(\ang{\bt{1}2}\squ{\bt{3}4}(-m^2-m_f^2-3u)+\ang{2\bt{3}}\squ{\bt{1}4}(3m_f^2+3m^2)\nonumber\\
&+&(\ang{\bt{13}}+\squ{\bt{13}})\bra{2}p\sket{4}(3m_f-m))\bra{2}p\sket{4}.
\end{eqnarray}
Again there is no consistent massless limit to be taken and a similar argument as in the scalar multiplet follows. Clearly, this issue requires further analysis.

\section{Discussion}
Using the recently developed formalism for massive scattering amplitudes, we have shown in this article that the chiral scalar multiplet and the non-chiral vector multiplet of massive supergravity both produce discontinuous scattering amplitudes in the $m\rightarrow 0$ limit, reproducing on-shell the field theory result of \cite{Deser:1977ur}. That said, scattering amplitudes themselves are not observables. Indeed, the primary significance of the original vDVZ discontinuity of massive gravity lies in the emperical fact that the light bending angle in massive gravity deviates from observation by a factor of $\frac{3}{4}$, provided the amplitudes are normalised to recover Newtonian gravity. This argument is obviously less relevant in the context of supergravity. However, our results establish the discontinuity at the amplitude level as an on-shell avatar of the St\"uckelberg decomposition. \\

Let's unpack this a little, beginning with our results from the supersymmetric case. In all the diagrams in this article, our conventions are that time moves upwards and external particles are considered outgoing. This results in a decided difference in the two amplitudes computed in the supersymmetric vector multiplet. The first that we computed in \eqref{eq:susy4vector1} describes the chiral multiplet. This amplitude does {\it not} have a discontinuity in the massless limit in that $\cl{M}_4[1^{+1/2},2^{-1},3^{-1/2},4^{+1}]\bigg|_{m\to 0}=\cl{M}^{3/2}_4[1^{+1/2},2^{-1},3^{-1/2},4^{+1}]\bigg|_{m=0}$. The next one that we computed, \eqref{eq:susy4vector2} describes a non-chiral multiplet. This amplitude does indeed exhibit a discontinuity in the massless limit, $\cl{M}_4[1^{-1/2},2^{-1},3^{+1/2},4^{+1}]\bigg|_{m\to0}=\frac13\cl{M}^{1/2}_4[1^{-1/2},2^{-1},3^{+1/2},4^{+1}]\bigg|_{m=0}$\footnote{The superscript on the amplitude here denotes the spin of the massless propagator.}. Guided by the fact that the field theory is chiral, we need only consider that piece. It is worth noting that if our only source of information came from on-shell amplitude methods we would necessarily have to include the non-chiral part, resulting in an ambiguity in the realization of discontinuity. An interesting  problem for the future would be to understand how to project out the chiral part of the amplitude. Moving on to the scalar multiplet, we find that 
       \begin{eqnarray*}\cl{M}_4^{I_1I_3}[\bt{1}^{1/2},\bt{2}^{0},\bt{3}^{1/2},\bt{4}^{0}]\bigg\vert_{m\to0}&=&\cl{M}^{3/2}_4{}^{I_1I_3}[\bt{1}^{1/2},\bt{2}^{0},\bt{3}^{1/2},\bt{4}^{0}]\bigg\vert_{m=0}\\ 
       &+& \frac{2}{3}\cl{M}^{1/2}_4{}^{I_1I_3}[\bt{1}^{1/2},\bt{2}^{0},\bt{3}^{1/2},\bt{4}^{0}]\bigg\vert_{m=0}, 
       \end{eqnarray*}
       manifesting the discontinuity anticipated from the field theory analysis. \\

Next, we attempted to see the discontinuity with supersymmetry broken, endowing the external particle species with distinct masses. Sadly, we were unable to express the relevant four-particle amplitudes in a form in which the massless limit can be cleanly taken. This is due to how the structure of the coupling functions in the three-particle amplitudes is determined. Specifically, the amplitudes can be organised such that they reveal a mass dependence of the order $\sim 1/m$ and is not finite in the massless limit while, for all of the massless limits to be well defined, we require a mass dependence of the form $1/(m+m_f-m_s)$. We remain unsatisfied with this puzzle but leave its resolution for future work. Another noteworthy point in the non-supersymmetric case is that the vector multiplet amplitude \eqref{eq:vector4massive} contains terms that correspond to both the chiral and non-chiral pieces as a result of the now massive fermion. \\

The on-shell technology developed in \cite{Arkani-Hamed:2017jhn} for massive particle scattering is both conceptually and computationally powerful, but like any new technology, it requires extensive beta-testing to iron out any bugs. This article details one such test. We set out to give an on-shell derivation of the spin-$\frac{3}{2}$ analogue of the famous vDVZ discontinuity of massive gravity, expecting the calculation to be clean and unambiguous. We were met with several subtlties, some of which we were able to resolve and some, like how to treat symmetry breaking and chirality, we remain puzzled by. Clearly though, there is still much work that remains to be done.

\section*{Acknowledgements}
We would like to thank the anonymous referee of one of our previous papers for pointing us to this problem. JM is supported by the NRF of South Africa under grant CSUR 114599. DB and NM is supported by funding from the DST/NRF SARChI in Physical Cosmology. Any opinion, finding and conclusion or recommendation expressed in this material is that of the authors and the NRF does not accept any liability in this regard.

\appendix
\section{Conventions and Notation} \label{sec:Conv}
In the interests of self-containment, we collect here some of our conventions for computing massive amplitudes that we use in the main text. Our convention for massive spinors can be summarized as:
\begin{itemize}
\item If no up-down massive spinor index pair is explicit on adjacent massive spinors they are considered to be contracted. 
\item For spinor indices we use the lower case Latin alphabet and for little-group indices we use the upper case Latin alphabet. 
\item  Epsilon conventions and helicity basis:
    \begin{eqnarray*}
       \epsilon_{ab} = \begin{bmatrix} 0 & -1 \\ 1 & 0 \end{bmatrix}\,,\quad
       \epsilon^{ab} = \begin{bmatrix} 0 & 1 \\ -1 & 0 \end{bmatrix}\,,\quad
       \zeta^{-I}= \begin{bmatrix} 1\\0\end{bmatrix}\,,\quad
        \zeta^{+I}= \begin{bmatrix} 0\\1\end{bmatrix} 
    \end{eqnarray*}
\item  Mandelstam variables:
    \begin{eqnarray*}
        s=-(p_1 + p_4)^2\,,\quad
        u=-(p_1 + p_3)^2\,,\quad
        t=-(p_1 + p_2)^2
    \end{eqnarray*}
\item Massless spinor relations:
    \begin{eqnarray*}
       p^{a\db} = -\ket{p}^{a}\sbra{p}^{\db}\,,\quad
       \ket{-p} =-\ket{p}\,,\quad
        \sket{-p} =\sket{p}
    \end{eqnarray*}
\item  Massive spinor relations:
    \begin{eqnarray*}
       p^{a\db} &=& \ket{\bt{p}}^{aI}{}_{I}\sbra{\bt{p}}^{\db} = -\ket{p}^a\sbra{p}^{\db}-\ket{\eta_p}^a\sbra{\eta_p}^{\db}\,,\quad
       p^2=\mathrm{det}(p^{a\db}) = -m_p^2\,,\\
       \ang{p\eta_p} &=& \squ{p\eta_p} = m_p \quad
       \bra{\bt{p}}_a^Ip^{a\db} = m \sbra{\bt{p}}^{\db I}\,,\quad
       \sbra{\bt{p}}^{\da I}p_{\da b} = -m \bra{\bt{p}}_{b}^{ I}\,,\\
       \ang{i|pp|j} &=& -m_p^2 \ang{ij}\,,\quad
       {}^I\ang{\bt{p}\bt{p}}^J = m \epsilon^{IJ}\quad
        {}_I\squ{\bt{p}\bt{p}}_J = -m \epsilon_{IJ}\,,\\
        \ang{i\bt{p}}^I{}_I\ang{\bt{p}j} &=& m\ang{ij}\,,\quad
        \squ{i\bt{p}}^I{}_I\squ{\bt{p}j} = m\squ{ij}\,,\quad
        \ket{\bt{p}}^{I}{}_{J}\sbra{\bt{p}} = -\ket{\bt{p}}_{I}{}^{J}\sbra{\bt{p}}
    \end{eqnarray*}
\end{itemize}

\section{Some Miscellaneous Identities}\label{app:ident}

\begin{itemize}
	\item Simplification of the general scalar multiplet amplitude where we symmetrize over the massive relevant massive indices
		\begin{equation}
			\begin{split}
			\ang{\bt{p}^{\{I}\bt{p}^{J\}}}
			&=m\epsilon^{\{IJ\}}=0,\\
			\bra{\bt{1}}p_1p\ket{\bt{p}}
			&= -m_f m \squ{\bt{1}\bt{p}},\\
			\bra{\bt{p}}p_1p\ket{\bt{1}}
			&=m m_f \squ{\bt{1}\bt{p}} - (m^2+m_f^2-m_s^2)\ang{\bt{1}\bt{p}},\\
			\bra{\bt{p}}p_1p\ket{\bt{p}}
			&=-m \bra{\bt{p}}p_1\sket{\bt{p}}.
			\end{split}
		\end{equation} 
	
	\item Here are some identities commonly used in the scalar multiplet amplitudes. We keep the masses of the three particles distinct for clarity. For the terms in the all massive three-particle amplitude that are not listed here, simply exchange $1\rightleftarrows3$ or where relevant complex conjugate.
		\begin{equation}
\begin{split}
\bra{\bt{p}^I}p_1\sket{\bt{p}^J}\sbra{\bt{p}_J}p_3\sket{\bt{p}_I} 
&= -m^2(2p_1\cdot p_3)\\
\bra{\bt{p}^I}p_1pp_3\sket{\bt{p}_I}
&= m^2(2p_1\cdot p_3)+(2p\cdot p_1)(2p\cdot p_3)\\
\bra{\bt{1}}pp_3pp_1p\sket{\bt{3}}
&= m^2\bra{\bt{1}}pp_3p_1\sket{\bt{3}} + m^2 m_f \ang{\bt{1}\bt{3}}(2p\cdot p_1) +\bra{\bt{1}}p\sket{\bt{3}}(2p\cdot p_1)(2p\cdot p_3)\\
\bra{\bt{1}}pp_3pp_1\ket{\bt{3}}
&=m^2\bra{\bt{1}}p_3p_1\ket{\bt{3}} - m_f \bra{\bt{3}}p\sket{\bt{1}}(2p\cdot p_3) + \ang{\bt{13}}(2p\cdot p_3)^2 \\				
\bra{\bt{1}}pp_3p_1p\ket{\bt{3}}
&=m^2\bra{\bt{1}}p_3p_1\ket{\bt{3}} - m_f\bra{\bt{1}}p\sket{\bt{3}}(2p\cdot p_1) + \ang{\bt{13}}(2p\cdot p_1)^2 - m_f \bra{\bt{3}}p\sket{\bt{1}}(2p.p_1)\\				
\bra{\bt{1}}p_3pp_1p\ket{\bt{3}}
&=-m^2\bra{\bt{1}}p_3p_1\ket{\bt{3}} +m_f\bra{\bt{1}}p\sket{\bt{3}}(2p\cdot p_1) - \ang{\bt{13}}(2p\cdot p_1)^2\\				
\bra{\bt{1}}p_3p_1p\sket{\bt{3}}
&=-m_f^2 \bra{\bt{3}}p\sket{\bt{1}} - m_f \squ{\bt{13}}(2p\cdot p_3) + \bra{\bt{1}}p\sket{\bt{3}}(2p_1\cdot p_3)\\
\bra{\bt{1}}p_3pp_1\sket{\bt{3}}
&=m_f^2\bra{\bt{3}}p\sket{\bt{1}} - m_f \ang{\bt{13}}(2p\cdot p_1) - \bra{\bt{1}}p\sket{\bt{3}}(2p_1\cdot p_3) + m_f \squ{\bt{13}}(2p\cdot p_3)\\
\bra{\bt{1}}pp_3p_1\sket{\bt{3}}
&=-m_f^2\bra{\bt{3}}p\sket{\bt{1}} + m_f \ang{\bt{13}}(2p\cdot p_1) + \bra{\bt{1}}p\sket{\bt{3}}(2p_1\cdot p_3)
\end{split}
\end{equation}

\end{itemize}

\bibliographystyle{JHEP}
\bibliography{mainbib}

\end{document}